\documentclass[conference]{IEEEtran}
\IEEEoverridecommandlockouts
\usepackage{cite}
\usepackage{amsmath,amssymb,amsfonts}
\usepackage{algorithmic}
\usepackage{graphicx}
\usepackage{textcomp}
\usepackage{blindtext}
\usepackage{float} 
\usepackage{subfigure}
\usepackage{xcolor}
\usepackage{wrapfig}
\usepackage{makecell}

\def\BibTeX{{\rm B\kern-.05em{\sc i\kern-.025em b}\kern-.08em
		T\kern-.1667em\lower.7ex\hbox{E}\kern-.125emX}}
\makeatletter

\begin{document}
	\title{Effect of Variable Physical Numerologies on Link-Level Performance of 5G NR V2X}
	\IEEEpeerreviewmaketitle
	\author{\IEEEauthorblockN{Donglin Wang\IEEEauthorrefmark{2}, Oneza Saraci\IEEEauthorrefmark{2}, Raja R.Sattiraju, Qiuheng Zhou\IEEEauthorrefmark{1}, and Hans D. Schotten\IEEEauthorrefmark{2}\IEEEauthorrefmark{1}}
		\IEEEauthorblockA{\textit{\IEEEauthorrefmark{2}University of Kaiserslautern, Kaiserslautern, Germany} \\
		$\{$dwang,sattiraju,schotten$\}$@eit.uni-kl.de \\
		$\{$saraci$\}$@rhrk.uni-kl.de}
		\IEEEauthorblockA{\textit{\IEEEauthorrefmark{1}German Research Center for Artificial Intelligence (DFKI GmbH), Kaiserslautern, Germany} \\
		$\{$qiuheng.zhou,schotten$\}$@dfki.de}
	}
	
	\maketitle
\begin{abstract}
With technology and societal development, the 5th generation wireless communication (5G) contributes significantly to different societies like industries or academies. Vehicle-to-Everything (V2X) communication technology has been one of the leading services for 5G which has been applied in vehicles. It’s used to exchange their status information with other traffic and traffic participants to increase traffic safety and efficiency. Cellular-V2X (C-V2X) is one of the emerging technologies to enable V2X communications. The first Long-Term Evolution (LTE) based C-V2X was released on the 3rd Generation Partnership Project (3GPP) standard. 3GPP is working towards the development of New Radio (NR) systems that it’s called 5G NR V2X. One single numerology in LTE cannot satisfy most performance requirements because of the variety of deployment options and scenarios. For this reason, in order to meet the diverse requirements, the 5G NR Physical Layer (PHY) is designed to provide a highly flexible framework. Scalable Orthogonal Frequency-Division Multiplexing (OFDM) numerologies make flexibility possible. The term numerology refers to the PHY waveform parametrization and allows different Subcarrier Spacings (SCSs), symbols, and slot duration. This paper implements the Link-Level (LL) simulations of LTE C-V2X communication and 5G NR V2X communication where simulation results are used to compare similarities and differences between LTE and 5G NR. We detect the effect of variable PHY Numerologies of 5G NR on the LL performance of V2X. The simulation results show that the performance of 5G NR improved by using variable numerologies.  

\end{abstract}


\section{Introduction}

With the development of the transportation system in the automotive industry, lots of innovative changes in this area  have been shown up over the past decade. Nowadays, Vehicles can meet high requirements and get ready for considering of serious autonomous driving. The 3GPP Release 14 (Rel.14) specification was first published Vehicular-to-everything (V2X) specification which is seen as an important technology to provide entire surrounding situation information with low latency and high reliability for the vehicle by exchanging information with other traffic, roadside units, and pedestrians [1]. There are two technologies emerged for enabling V2X communication. The first one is the Dedicated Short Range Communications (DSRC) standard which is referred to as ITS-G5 in Europe. There are plenty of research from last two decades on ITS-G5 which is a mature technology [2]. Another technology is the LTE-based C-V2X which provides communication services in vehicular scenarios in 2017 [1]. However, with more high quality of service requirements, LTE C-V2X is not able to meet the requirements in many V2X applications. So in the 3GPP Release 15 (Rel.15), the NR technology was released and some more enhancements to Rel.14 C-V2X are added and taken as a complementary solution [3]. 5G NR-based V2X was released in Release 16 (Rel.16) in 2019 which focuses more research on direct vehicular communication through the sidelink PC5 interface [4]. Many works have evaluated the performance of C-V2X communication. The system-level performance for transmission mode 3 of direct LTE C-V2X communication has been done in [5][6]. In [7], it's shown that LTE C-V2X gets a superior reliability performance over DSRC through the PHY layer performance comparison between ITS-G5 and C-V2X communications by using different channel Vehicular models. 

In [8], the studies on 5G NR V2X have started. [9][10] are tutorials on 5G NR V2X communications. In work [11], in terms of network, architecture, security protocol enhancements in NR V2X in 3GPP Rel.16 are designed. In [12], Patriciello published 5G NR numerologies and their impact on end-to-end latency in 2018. Because 5G can support different high carrier frequencies, the numerology effect on 5G 28 GHz communication system performance was detected in [13]. In [14], the effect of 5G NR numerologies on V2X communications has been evaluated, but just for 15 kHz, 60 kHz, and 120 kHz. However, these available publications for 5G NR technology are about the 3GPP standard overviews, less of them discussed the comprehensive implementations or simulation results to check the performance of 5G NR V2X communication for all different numerologies. Less of these works are done in the performance comparison between LTE and 5G NR. 

In this work, we apply the latest 5G NR technology for V2X communication and compare the LL performance between LTE C-V2X and 5G NR V2X communications. 
The rest of the paper is organized as follows: in Section II, we have the PHY comparison of LTE and 5G NR. Next in section III, the simulation methodology and simulation settings are given. Section IV reports the simulation results and analysis. Finally, Section V draws the conclusion and our future work plans. 

\section{Two technologies comparison}
In this part, brief overviews regarding the PHY layer for LTE C-V2X and 5G NR V2X are provided. In the following Table I, there are different comparisons between LTE and 5G NR of these PHY parameters. 
\subsection{LTE-based V2X communication}

\begin{table*}[htbp]
\caption{Comparison between LTE C-V2X and 5G NR V2X}
\begin{center}
	\begin{tabular}{|p{6cm}<{\centering}|p{4cm}<{\centering}|p{6cm}<{\centering}|}
		\hline 
		Parameters&	                     LTE&           5G NR\\
		\hline
		
		Carrier frequency&		        up to 6 GHz&    \thead{FR1: 410-7125\\ FR2: 24250-52600 (MHz)}  \\  
		\hline
		Supported channel bandwidths&	Max.20 MHz&     \thead{FR1:  5, 10, 15, 20, 25, 30, 40, 50, 60, 80, 90, 100 \\ FR2: 50, 100, 200, 400 (MHz)}\\  
		\hline
		SCS&                            15 KHz&          15, 30, 60, 120, 240  (KHz) \\  
		\hline
		Slot length&                    7 symbols&      14 symbols\\
		\hline
		Max number of subscarriers&		1200&	        3300\\  
		\hline
		Modulation&	                	SC-OFDM&        OFDM \\  
		\hline
		Channel coding&              \thead{ Data: Turo coding\\ Control: convolution coding} &    \thead{Data: LDPC coding \\ Control: Polar coding} \\
		\hline
		Latency&                        10 ms&          1 ms \\
		\hline
		Communication types&            Broadcast&      Broadcast, groupcast, unicast \\
		\hline
		Retransmission&                Blind retransmission& Blind and feedback-based retransmission \\
		\hline
	   Sidelink	resource allocation modes&    Mode 3 and mode 4&   Mode 1 and Mode 2    \\
		\hline
	\end{tabular}
\end{center}
\end{table*}
\begin{figure}[htbp]
	\centering
	\includegraphics[width=\linewidth]{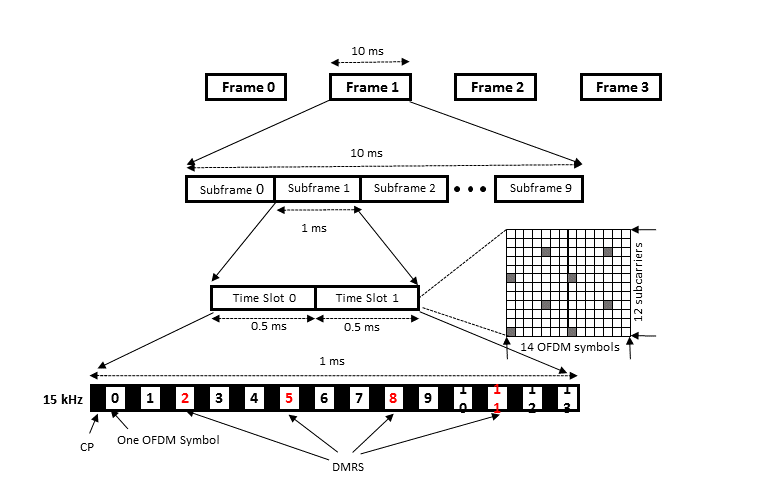}
	\caption{LTE time domain frame structure}
	\label{fig}
\end{figure} 

As introduced before, LTE C-V2X is the underlying technology that provides communication services in vehicular scenarios, which can optimize connected vehicles and transportation conditions. As Fig. 1 shows, in line with the LTE specification, a radio frame is the largest unit of the LTE C-V2X frame structure (10 ms), and each frame is divided into ten equally sized subframes (1 ms). Moreover, each subframe is further divided into two equally sized time slots where each slot is 0.5 ms, so there is a total of 20 slots in a radio frame. Each slot is further divided into OFDM symbols. The number of OFDM symbols in LTE is 14 symbols per subframe and 140 symbols per radio frame in the time domain. Symbols 2, 5, 8, and 11 are used for transmitting Demodulation Reference Signal (DMRS) marked in red in Fig. 1 utilized for frequency correction and channel estimation. the actual data is carried by using the left 10 symbols. The PHY layer of the C-V2X communication is designed as identical to the LTE uplink [1]. For the resource scheduling assignment, the control data and the shared data  of uplink resources are transmitted in a single subframe over adjacent Physical Resource Blocks (PRBs). The number of PRBs in the frequency domain is dependent on the bandwidth. Taking the 10 MHz bandwidth into consideration, there is a total of 50 PRBs. Each PRB has 12 sub-carriers and each sub-carrier is 15 kHz in the frequency domain [15]. For direct C-V2X communication where traffic Users (UEs) can directly transmit data packets to other traffic participants in their proximity, that's the sidelink transmission, the physical resources including subframes and resource blocks connected with a given pool are partitioned into a sequence of periodic repeating PRBs known as Physical Sidelink Shared Channel (PSSCH) periods [1]. In a PSSCH period, for the control and real transmission data, separate subframe pools and resource block pools are used.
The dynamic transmission properties of the PSSCH are described by the Sidelink Control Messages (SCIs) which is carried by the Physical Sidelink Control Channel (PSCCH). The receiver searches all configured PSSCH resource pools for SCI transmissions of interest to it [1]. Another important concept in LTE C-V2X is the Cyclic Prefix (CP). LTE C-V2X has two types of CP: normal CP and extended CP. However, LTE-V2X PC5 supports only the normal CP. A CP has been created to  preserve the orthogonality of the subcarriers and prevent Intersymbol Interference (ISI) between successive OFDM symbols [16]. CP is the same repetition of the last part of the OFDM symbol attached before the OFDM symbol, as shown in Fig. 1. 

\subsection{5G NR based-V2X communication}
5G NR V2X is expected and designed to meet various requirements of different use cases over LTE C-V2X. 5G NR has multiple fresh characteristics and functionalities which essentially contribute to improving the data rate, reducing latency, and improving the spectral efficiency of the V2X communication systems.
\subsubsection{Communication types and new sidelink feedback channel}
with a view to supporting a wide variety of applications, NR V2X supports three types of transmissions: broadcast, gourpcast, and unicast [4]. 
There is a new channel Physical Sidelink Feedback Channel (PSFCH) in 5G NR V2X for sending Hybrid Automatic Repeat Request (HARQ) feedback. However, PSFCH is only required for the unicast and groupcast transmission modes which are not considered in LTE C-V2X [4]. 
\subsubsection{Radio access}
To cope with various deployment scenarios, the 5G NR radio access technology support wide bandwidth in wide frequency ranges including sub 6 GHz bands and millimeter Wave (mmWave) bands (within two possible ranges) and channel bandwidths, as shown in Table I. Frequency 1 (FR1) carrier frequency is from 410 to 7125 (MHz) and channel bandwidths are 5, 10, 15, 20, 25, 30, 40, 50, 60, 80, 90, and 100 (MHz). Frequency 2 (FR2) carrier frequency range is from 24250 to 52600 (MHz) and channel bandwidths are 50, 100, 200, and 400 (MHz) [17].  
\subsubsection{Resource allocation modes}
Similarly to LTE C-V2X, which uses transmission mode 3 and mode 4 for LTE C-V2X sidelink communication, mode 3 is using Base Station (BS) to schedule the transmission resources for traffic UEs; in mode 4, vehicles select their resource from the sidelink resource pool autonomously[1]. In 5G NR, there are also two resource allocation modes for 5G NR V2X sidelink communication. One is centralized (transmission mode 1): the 5G NR BS (gNB) schedules radio resources for traffic UEs for 5G NR V2X sidelink communication where traffic UEs are in the coverage of the gNB.  
And another is distributed (transmission mode 2): the traffic UEs autonomously determine sidelink transmission resources within sidelink resources. The sidelink resources are configured by gNB or pre-configured by the network where traffic UEs are not necessary under the coverage of the gNB or the network [4].

\begin{table*}[htbp]
\caption{Multiple Numerologies in NR}
\begin{center}
	\begin{tabular}{|p{1.5cm}<{\centering}|p{1.5cm}<{\centering}|p{2cm}<{\centering}|p{1.5cm}<{\centering}|p{1.5cm}<{\centering}|p{2cm}<{\centering}|p{2cm}<{\centering}|}
		\hline 
		CP&	        SCS [KHz]& $\#$subframes per radio frame&   $\#$slots per subframes&  slot duration(ms)& $\#$OFDM symbols per slot& 	 Applicable frequency range \\
		\hline
		normal&		15&	       10&	 1& 	1& 14& FR1\\  
		\hline
		normal&		30&     	10&	 2&	    0.5& 14& FR1\\  
		\hline
		normal&		60&	        10&	 4&	    0.25& 14& FR1 and FR2\\  
		\hline
		normal&		120&        10&  8&      0.125& 14& FR2\\  
		\hline
	\end{tabular}
\end{center}
\end{table*}

\subsubsection{Numerologies}
In 5G NR, to allow for such flexibility, NR uses a flexible frame structure, with different SCSs. The SCS is the distance between the centers of two consecutive subcarriers, this is referred to as "multiple numerologies". The specifications have introduced numerology as parameter $\mu$. The design parameter $\mu$ relies on multiple factors together with the type of deployments, mobility, performance, and implementation, etc. The SCS is calculated with the formula: $ \Delta f = 2^\mu * 15 $ where the integer-valued $\mu$ can be optimized for different scenarios. In 3GPP 5G NR Rel.15, $\mu$ takes the values from 0 to 4 [3]. Numerology is effectively an indication of the SCS. So SCS of 15, 30, 60, 120, and 240 (KHz) are supported in 5G NR. However, 5G NR V2X supports up to 120 kHz in FR2. 240 kHz SCS is not supported. The reason for multiple SCS for NR is that 5G must be deployed in lower frequencies as well as in high frequencies. For FR1, only SCS of 15 kHz, 30 kHz, or 60 kHz is used, while for FR2, which refers to mmWave frequencies, only 60 kHz or 120 kHz can be used for data transmission. 

As shown in Fig. 2, 5G NR is designed with a time-domain structure that contains frames of 10 ms. Each frame is again divided into ten equal-sized subframes of 1 ms as LTE. Then this subframe is again divided into time slots consisting of 14 OFDM symbols. The default number of the OFDM symbols is 14 per slot. From the frequency domain of 5G NR, one RB is always 12 subcarriers. As the SCS depends on the numerology, when we change the numerology, the SCS changes, and therefore this affects the size of PRB. For example, When $\mu$ is 0, then SCS is 15 kHz, the PRB size is $12*15=180$ KHz, or if $\mu$ is 2, SCS is 30 kHz, then PRB size is $12*30=360$ kHz.

As shown in Table II, for SCS of 15 kHz and normal CP, a subframe contains only one slot, and a radio frame has 10 slots. For SCS of 30 kHz and normal CP, a subframe contains 2 slots, which means a radio frame contains 20 slots, and so on. it's obvious when moving from 15 kHz to 30 kHz, the number of slots is double, which is to say that the number of symbols in the time domain also doubles. So, for the same subframe of 1 ms, the slot duration from SCS of 15 kHz to SCS of 30 kHz is reduced from 1 ms to 0.5 ms and more OFDM symbols are supported with a very small duration. The same procedure can be applied for SCS of 60 kHz and 120 kHz. 
The support of multiple numerologies and multiple SCSs is the most outstanding 5G NR new feature when compared to LTE.
\begin{figure}[htbp]
	\centering
	\includegraphics[width=\linewidth]{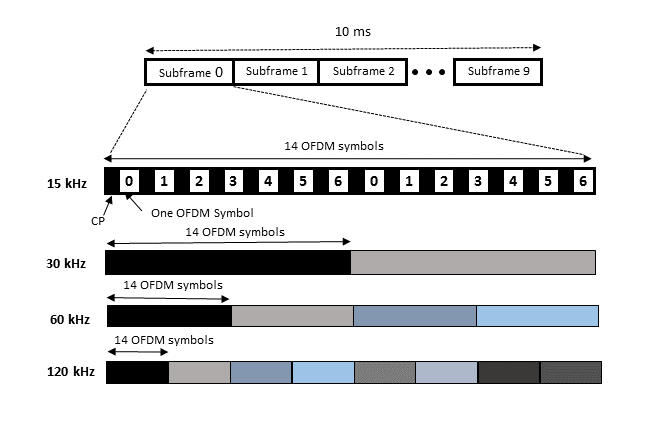}
	\caption{5G NR time domain frame structure}
	\label{fig}
\end{figure} 

\subsubsection{CP design in 5G NR}
The basic design of CP in NR is similar to LTE and same overhead as that in LTE as we mentioned in the previous chapter. 
CP design of 5G NR will depend on numerology. The CP length for different SCS can be calculated using the following formula based on 3GPP TS 38.211 [3].

\begin{equation}
C=7*2^\mu     \label{eq}
\end{equation}
In Eq.1, this is a conditional equation where $\mu$ is the numerology parameter.
\begin{equation}
K = \frac{T_s}{T_c} =64 \label{eq}
\end{equation}
In this equation, $k$ is a constant number. $T_s$ is the LTE basic time and $T_c$ is NR basic time [3]. 

The CP length for different SCSs can be calculated using the following formulas: 
\begin{equation} \label{eq}
      N^u_{CP,l}  =
\begin{cases}
    
    512k*2^{-\mu}     &\text{$extended \: CP$}\\
    144k*2^{-\mu}+16k &\text{$normal \:CP, l=0 \:or \:l= C$}\\
    144k*2^{-\mu}     &\text{$normal  \:CP, l \neq 0 \:and \: l \neq$}

\end{cases} 
\end{equation}
where $l$ is the symbol index here. The normal CP is specified for all SCSs. The extended CP is currently designed for the 60 kHz SCS. 
Then, the CP time duration can be calculated using the following formula:
\begin{equation}
T_{cp} = N_{cp}*T_c  \label{eq}
\end{equation}


All above, these new improvements and key technologies like resource allocation modes, numerology, and CP design have been designed to improve the reliability of V2X communications systems in 5G NR. 

\section{simulation methodology and simulation settings}
For the link simulation, it's necessary to build the entire transmission and receiving operations. We have implemented the LL simulator for IEEE 802.11p and LTE C-V2X in Python in our previous work where more simulator details can be found in [7]. The simulation pipeline is the same as the Section.III.B in [7]. In Fig. 3, the simulation processing procedure is shown. Here we give short introductions about each processing step. 

\subsubsection{Control channel processing}
In this step, a new SCI message is created which includes the MCS value, resource indication value, group destination identity, etc. The created SCI is a binary message which is encoded by using a convolutional encoder followed by rate matching, interleaving, and a 16-bit Cyclic Redundancy Check (CRC) attached to the encoded message. Once we have the binary codes, the next step is to process PSCCH scrambling. There are 240 PSCCH-generated symbols and are cyclically shifted with a random value chosen from the set [0, 3, 6, 9] to reduce the effect of interference. Then, 4 DMRS symbols are generated and mapped to the remaining 4-time domain symbols [2, 5, 8, 11].
\subsubsection{Shared channel processing}
In this procedure, a lot of processing steps are included, such as the type-24A CRC calculation, turbo encoding, rate matching, code block concatenation, interleaving, etc. The Quadrature Phase Shift Keying (QPSK) or 16QAM is used here to modulate the generated code. In order to generate the data symbols, the Discrete Fourier Transformation (DFT) is followed by means of transform precoding. it's similar to the control channel, data symbols are transmitted by adding DMRS symbols in a share channel subframe. Later, all symbols are mapped to the sidelink resource pool grid. The SC-OFDM is used to modulate symbols to create the time domain waveform which is filtered by a channel with an added Additive White Gaussian Noise (AWGN) noise. 
\subsubsection{Receiver operations}
In the receiving processing step, the receiver tries to perform a blind decoding of all available control channel subframes and DMRS CP. For each candidate, the channel is estimated using the DMRS symbols followed by equalization, PSCCH decoding, and finally SCI message decoding. If the CRC check is passed, then the control message is marked as decoded. The same procedures are repeated for decoding the data message, channel estimation, equalization, PSSCH decoding to get the CRC check and the decoded data. 
\begin{figure}[htbp]
	\includegraphics[width=\linewidth]{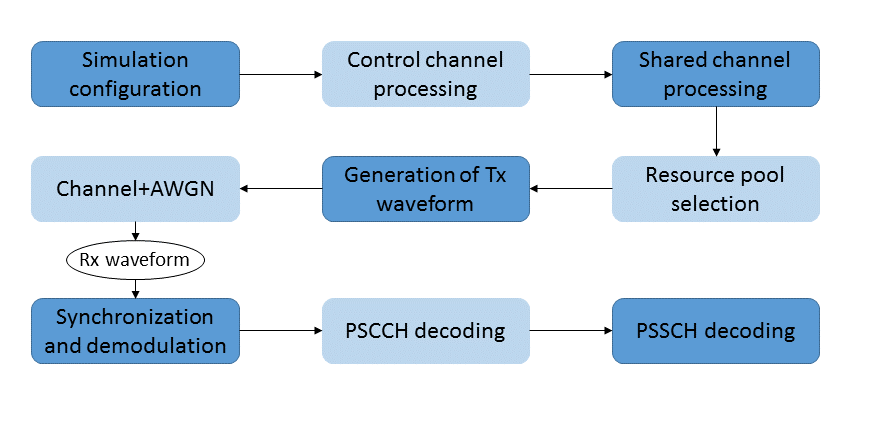}
	\caption{C-V2X LL transmitting and receiving chart [20]}
	\label{fig}
\end{figure} 

In this work, we use the existing LL simulator for 5G NR V2X communication by adding more 5G NR features such as numerologies, variable SCSs, CP, and increased bandwidth. 
We calculate the Block Error Rate (BLER) of the control and shared channels of Rel.14 with features of 5G NR V2X. BLER is a ratio of the number of error blocks to the complete number of transmitted blocks. Signal to Noise Ratio (SNR) is defined as the ratio of signal power to noise power, which is used to compare the desired signal to the background noise. In this work, the relationship between BLER and SNR is investigated to check the LL performance of 5G NR V2X.
In this simulator, some more simulation parameters can be found in Tab.III. For example, the UE speed is high which is 120 km/s. The ITU-Extended Vehicular A (EVA) channel model is used for V2V communication [18]. 
\begin{table}[htbp]
\caption{Simulation parameters}
\begin{center}
\begin{tabular}{ |p{4cm}|p{2cm}|  }
 \hline
 \textbf{Parameters} &  \textbf{Value} \\
 \hline
 Carrier frequency& 5.9 GHz\\
 \hline
 Bandwidths& 20 MHz\\
 \hline
 Packet size& 225 bytes\\
 \hline
 Number of Subframes& 50 \\
 \hline
 Data NPRBs& 8\\
 \hline
 TBS size& 1800\\
 \hline
 Data modulation& 16QAM\\
 \hline
 SNR range & 11-15 dB\\
 \hline
 MCS index& 13\\
 \hline
 Channel model&EVA\\
 \hline 
 UE speed & 120km/h \\
 \hline
 Max Doppler shift&180\\
 \hline 
 Doppler Model &Jakes [19] \\
 \hline
 number of Receiver antennas & 2\\
 \hline 
 PSCCH symbols & 240 \\
 \hline 
 DMRS & [2,5,8,11] \\
 \hline
\end{tabular}
\end{center}
\end{table}
\section{simulation results}
In this simulation work, several simulations are performed intending to see how numerology affects the system’s performance. As a reference, we get the LTE C-V2X with a fixed SCS of 15 kHz from our previous work as shown in Fig. 3 [7]. To bring the 5G NR features, we change the SCSs and present the simulation results for each case, as shown in Fig. 4, the results presented below are generated from simulations for $\mu = 0,1,2,3$ (SCSs of 15, 30, 60, and 120 (KHz)) of 5G NR.

The performance of the system counts on the value of BLER. Lower BLER means a better performance of the 5G NR and vice-versa. As shown in Fig. 4, for SCS of 15 kHz, the results of LTE C-V2X and 5G NR V2X performance are almost the same, with only a slight gain in 5G NR. This is what we expected, as for a channel bandwidth of 20 MHz that is used in both cases, in LTE C-V2X and 5G NR V2X, the number of Fast Fourier Transform (FFT) is the same, as same as the number of OFDM symbols per subframe which is 14 OFDM symbols in both cases. The only thing that changes is that in LTE C-V2X number of PRBs is 100, while in 5G NR, the number of PRBs is 106. This is a slight difference that doesn't affect too much the overall performance, which explains why the 15 kHz SCS is so similar to LTE C-V2X and 5G NR V2X.

Let's look at the results from Fig. 5, the system undergoes a slight improvement by increasing the SCS from 5G NR 15 kHz to 5G NR 30 KHz. Let's take the example of SNR 13 dB; for SCS 15 KHz, the BLER is 0.16, while for SCS of 30 kHz, the BLER is 0.1. We see that the BLER decreases, and the same happens for SNR 12 dB, SNR 14 dB, and SNR 15 dB, which makes the SCS 30 kHz the case with the best performance from all the simulations that were generated. This is because the further apart the sub-carriers are, the less likely it is that the frequency shift due to the Doppler effect on any sub-carrier will interfere with adjacent sub-carriers. Therefore, the system is more robust to Doppler effects. 

Moreover, when we increase the SCS size from 30 kHz to 60 kHz and then to 120 kHz, we don't gain in performance for small SNRs, but for larger SNRs, we see an improvement in the performance. 
The effect that occurs in the time domain by increasing the SCS must be considered here, i.e. reducing the OFDM symbol duration and thus also the slot duration and CP. For example, because the symbol duration and CP are four times shorter at SCS at 60 kHz than at 15 kHz, the system is less protected because CP is not enough to protect the signal from echoes.

Also, another aspect to look into is SNR's effect on BLER. As shown in Fig. 5, with the increase of SNR, the BLER decreases. The decrease of BLER while increasing the SNR happens for all cases of SCS. For example, in SCS 30 KHz, when we increase the SNR from SNR 11 dB to SNR 13 dB, the BLER goes from 0.36 to 0.1. Lower BLER means better performance, which means that the system's performance increases when SNR increases. Moreover, from Fig. 4 can be seen that for lower SCS (LTE C-V2X 15 kHz, 5G NR 15 kHz, and 5G NR 30 kHz), the performance is better compared with wider SCS (5G NR 60 KHz and 5G NR 120 kHz). This can be explained that with the increase of SCS, the interferences on the system also increase and affect the overall performance of the system.


\begin{figure}[htbp]
	\centering
	\includegraphics[width=\linewidth]{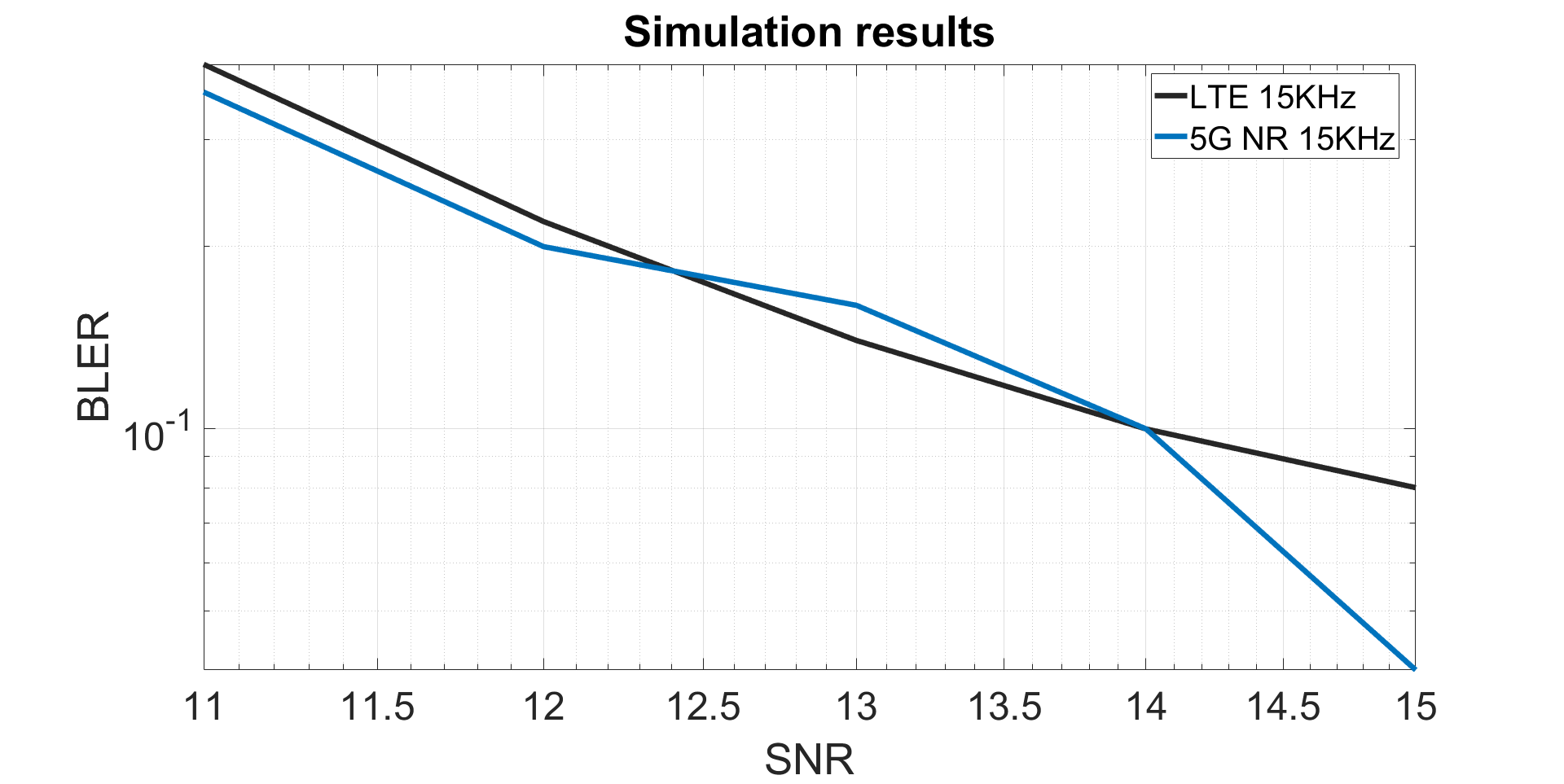}
	\caption{Comparsion between LTE and 5G NR for SCS of 15 KHz}
	\label{fig}
\end{figure} 

\begin{figure}[htbp]
	\centering
	\includegraphics[width=\linewidth]{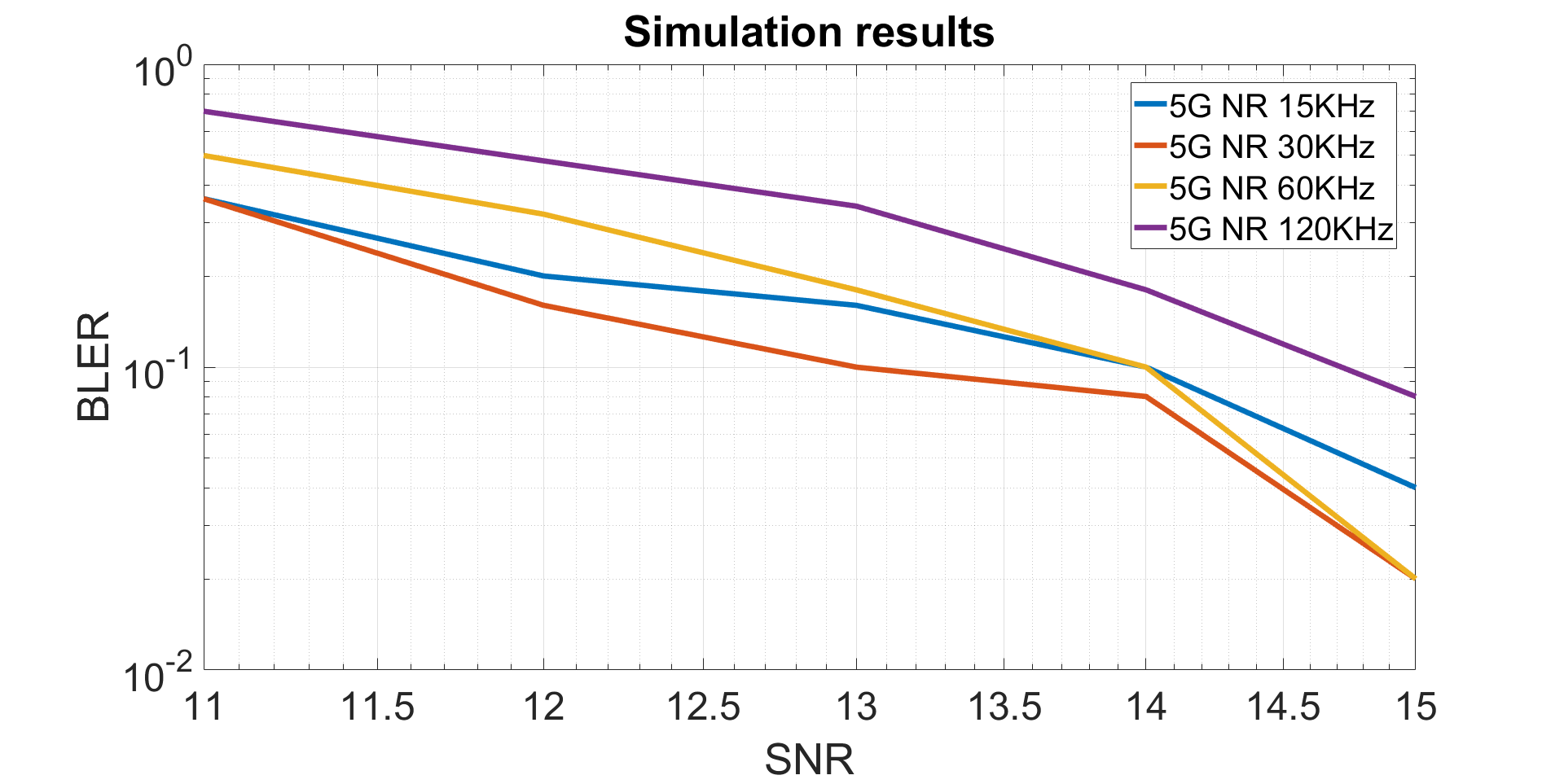}
	\caption{Simulation results for multiple SCSs of 5G NR}
	\label{fig}
\end{figure} 

\section{conclusion}
In this work, we introduced LTE C-V2X communication and 5G NR V2X communication. The PHY features of LTE and 5G NR are compared to get a better understanding of the differences between these two advanced technologies. Moreover, scalable numerology variables SCSs, CP, and waveforms are the most important features of the 5G NR technology we evaluated. To check the new advancements of the 5G NR, We implemented these 5G NR new features to our previous sidelink LTE C-V2X LL simulator and evaluated the PHY performance of 5G NR.  

We evaluated the PHY performance of the 5G NR by using the BLER and the SNR values. From the simulation results, it's obvious to tell in all SCS cases, with increasing SNR values, or with decreasing BLER values, the performance of the 5G NR improves. For SCS of 15 kHz in LTE C-V2X and 5G NR V2X, there are showing almost the same performance results. But for the SCS of 30 kHz, we have the most gain performance compared with other SCSs we implemented. From the simulation results,
we can also find more characteristics of different numerology and SCSs. It’s essential to acknowledge that numerology depends a lot on the deployment scenarios, which will affect the overall performance of the C-V2X communication system. More simulation results regarding 5G NR like the effect of retransmission on the performance of 5G NR will follow up.

\end{document}